\documentclass[onecolumn,prd,nofootinbib,aps,floats,floatfix,amsmath,amssymb,secnumarabic,preprintnumbers,showpacs,notitlepage]{revtex4-1} %
\usepackage[final]{graphicx}
\usepackage{ulem}
\usepackage{amsmath}

\def\beq{\begin{equation}}
\def\eeq{\end{equation}}
\def\bea{\begin{eqnarray}}
\def\eea{\end{eqnarray}}
\def\nn{\nonumber}

\def\roughly#1{\mathrel{\raise.3ex\hbox
{$#1$\kern-.75em\lower1ex\hbox{$\sim$}}}}
\def\lsim{\roughly<}
\def\gsim{\roughly>}
\def\s{\sqrt{2}}

\def\bs{B^0_s}
\def\bsbar{{\bar B}^0_s}

\def\bs{B^0_s}
\def\bsbar{{\bar B}^0_s}
\def\bsee{b \to s e^+ e^-}
\def\bsmumu{b \to s \mu^+ \mu^-}

\def\bsee{b \to s e^+ e^-}
\def\bsll{b \to s \ell^+ \ell^-}
\def\bsnunubar{b \to s \nu {\bar\nu}}

\def \cB{{\cal B}}
\def \SM{{\rm SM}}
\def \NP{{\rm NP}}
\def \expt{{\rm expt}}

\begin{document}

\title{\boldmath New physics in $\bsee$?}
\preprint{UdeM-GPP-TH-19-267}
\author{Jacky Kumar}
\email{jacky.kumar@umontreal.ca}
\affiliation{Physique des Particules, Universit\'e de Montr\'eal, \\
C.P. 6128, succ. centre-ville, Montr\'eal, QC, Canada H3C 3J7}
\author{David London}\email{london@lps.umontreal.ca}
\affiliation{Physique des Particules, Universit\'e de Montr\'eal, \\
C.P. 6128, succ. centre-ville, Montr\'eal, QC, Canada H3C 3J7}

\begin{abstract}
At present, the measurements of some observables in $B \to K^*
\mu^+\mu^-$ and $B_s^0 \to \phi \mu^+ \mu^-$ decays, and of
$R_{K^{(*)}} \equiv \cB(B \to K^{(*)} \mu^+ \mu^-)/{\cal B}(B \to
K^{(*)} e^+ e^-)$, are in disagreement with the predictions of the
standard model. While most of these discrepancies can be removed with
the addition of new physics (NP) in $\bsmumu$, a difference of $\gsim
1.7 \sigma$ still remains in the measurement of $R_{K^*}$ at small
values of $q^2$, the dilepton invariant mass-squared. In the context
of a global fit, this is not a problem. However, it does raise the
question: if the true value of $R_{K^*}^{low}$ is near its measured
value, what is required to explain it? In this paper, we show that, if
one includes NP in $\bsee$, one can generate values for
$R_{K^*}^{low}$ that are within $\sim 1\sigma$ of its measured value.
Using a model-independent, effective-field-theory approach, we
construct many different possible NP scenarios. We also examine
specific models containing leptoquarks or a $Z'$ gauge boson. Here,
additional constraints from lepton-flavour-violating observables,
$\bs$-$\bsbar$ mixing and neutrino trident production must be taken
into account, but we still find a number of viable NP scenarios.  For
the various scenarios, we examine the predictions for $R_{K^{(*)}}$ in
other $q^2$ bins, as well as for the observable $Q_5 \equiv
P^{\prime\mu\mu}_5 - P^{\prime ee}_5$.
\end{abstract}

\maketitle

\section{Introduction}

At the present time, there are a number of measurements of $B$-decay
processes that are in disagreement with the predictions of the
standard model (SM). Two of these processes are governed by $\bsmumu$:
there are discrepancies with the SM in several observables in $B \to
K^* \mu^+\mu^-$ \cite{BK*mumuLHCb1, BK*mumuLHCb2, BK*mumuBelle,
  BK*mumuATLAS, BK*mumuCMS} and $B_s^0 \to \phi \mu^+ \mu^-$
\cite{BsphimumuLHCb1, BsphimumuLHCb2} decays. There are two other
observables that exhibit lepton-flavour-universality violation,
involving $\bsmumu$ and $\bsee$: $R_K \equiv \cB(B^+ \to K^+ \mu^+
\mu^-)/{\cal B}(B^+ \to K^+ e^+ e^-)$ \cite{RKexpt} and $R_{K^*}
\equiv {\cal B}(B^0 \to K^{*0} \mu^+ \mu^-)/\cB(B^0 \to K^{*0} e^+
e^-)$ \cite{RK*expt}. Combining the various $\bsll$ observables,
analyses have found that the net discrepancy with the SM is at the
level of 4-6$\sigma$ \cite{Capdevila:2017bsm, Altmannshofer:2017yso,
  DAmico:2017mtc, Hiller:2017bzc, Geng:2017svp, Ciuchini:2017mik,
  Celis:2017doq, Alok:2017sui}.

All observables involve $\bsmumu$.  For this reason, it is natural to
consider the possibility of new physics (NP) in this decay.  The
$\bsmumu$ transitions are defined via an effective Hamiltonian with
vector and axial vector operators:
\bea
H_{\rm eff} &=& - \frac{\alpha G_F}{\s \pi} V_{tb} V_{ts}^*
      \sum_{a = 9,10} ( C_a O_a + C'_a O'_a ) ~, \nn\\
O_{9(10)} &=& [ {\bar s} \gamma_\mu P_L b ] [ {\bar\mu} \gamma^\mu (\gamma_5) \mu ] ~,
\label{Heff}
\eea
where the $V_{ij}$ are elements of the Cabibbo-Kobayashi-Maskawa (CKM)
matrix and the primed operators are obtained by replacing $L$ with
$R$. The Wilson coefficients (WCs) include both the SM and NP
contributions: $C_X = C_{X,\SM} + C_{X,\NP}$.  It is found that, if
the values of the WCs obey one of two scenarios\footnote{These numbers
  are taken from Ref.~\cite{Alok:2017sui}.  Other analyses find
  similar results.} -- (i) $C_{9,\NP}^{\mu\mu} = -1.20 \pm 0.20$ or
(ii) $C_{9,\NP}^{\mu\mu} = -C_{10,\NP}^{\mu\mu} = -0.62 \pm 0.14$ --
the data can all be explained.

In fact, this is not entirely true. $R_{K^*}$ has been measured in two
different ranges of $q^2$, the dilepton invariant mass-squared
\cite{RK*expt}:
\bea
R_{K^*}^\expt &=&
0.660^{+0.110}_{-0.070}~{\rm (stat)} \pm 0.024~{\rm (syst)} ~,~~ 0.045 \le q^2 \le 1.1 ~{\rm GeV}^2 ~, \nn\\
R_{K^*}^\expt &=&
0.685^{+0.113}_{-0.069}~{\rm (stat)} \pm 0.047~{\rm (syst)} ~,~~ 1.1 \le q^2 \le 6.0 ~{\rm GeV}^2 ~.
\label{RK*meas}
\eea
We refer to these observables as $R_{K^*}^{low}$ and $R_{K^*}^{cen}$,
respectively. At low $q^2$, the mass difference between muons and
electrons is non-negligible \cite{RK*theory}, so that the SM predicts
$R_{K^*}^{low,\SM} \simeq 0.93$ \cite{flavio}. For central values of
$q^2$ (or larger), the prediction is $R_{K^*}^{cen,\SM} \simeq 1$. The
deviation from the SM is then $\sim 2.4\sigma$ ($R_{K^*}^{low}$) or
$\sim 2.5\sigma$ ($R_{K^*}^{cen}$).  Assuming NP is present in
$\bsmumu$, one can compute the predictions of scenarios (i) and (ii)
for the value of $R_{K^*}$ in each of the two $q^2$ bins. These are
\bea
{\rm (i)}~~C_{9,\NP}^{\mu\mu} = -1.20 \pm 0.20 &~:~& 
R_{K^*}^{low} = (0.89) ~ 0.89 ~~,~~~~ R_{K^*}^{cen} = (0.81) ~ 0.83 ~, \nn\\
{\rm (ii)}~~C_{9,\NP}^{\mu\mu} = -C_{10,\NP}^{\mu\mu} = -0.62 \pm 0.14 &~:~& 
R_{K^*}^{low} = (0.84)~0.85 ~~,~~~~ R_{K^*}^{cen} = (0.67) ~ 0.73 ~.
\label{bsmumuNP}
\eea
In each line above, the final number is the predicted value of the
observable for the best-fit value of the WCs in the given scenario.
The number to the left of it (in parentheses) is the smallest
predicted value of the observable within the $1\sigma$ (68\% C.L.)
range of the WCs. We see that the experimental value of
$R_{K^*}^{cen}$ can be accounted for [though scenario (ii) is better
  than scenario (i)]. On the other hand, the experimental value of
$R_{K^*}^{low}$ cannot -- both scenario predict considerably larger
values than what is observed.

Now, scenarios (i) and (ii) are the simplest solutions, in that only
one NP WC (or combination of WCs) is nonzero. However, one might
suspect that the problems with $R_{K^*}^{low}$ could be improved if
more than one WC were allowed to be nonzero. With this in mind, we
consider scenario (iii), in which $C_{9,\NP}^{\mu\mu}$ and
$C_{10,\NP}^{\mu\mu}$ are allowed to vary independently. The best-fit
values of the WCs, as well as the prediction for $R_{K^*}^{low}$, are
found to be
\beq
{\rm (iii)} ~~ C_{9,\NP}^{\mu\mu} = -1.10 \pm 0.20 ~,~~ C_{10,\NP}^{\mu\mu} = 0.28 \pm 0.17 : 
R_{K^*}^{low} = (0.85)~0.87 ~.
\label{scenario(iii)}
\eeq
(Note that the errors on the WCs are highly correlated.) The number in
parentheses is the smallest predicted value of $R_{K^*}^{low}$ within
the 68\% C.L.\ region in the space of $C_{9,\NP}^{\mu\mu}$ and
$C_{10,\NP}^{\mu\mu}$.  We see that the predicted value of
$R_{K^*}^{low}$ is not much different from that of scenarios (i) and
(ii). Evidently, NP in $C_{9,\NP}^{\mu\mu}$ and/or
$C_{10,\NP}^{\mu\mu}$ does not lead to a sizeable effect on
$R_{K^*}^{low}$.

What about if other WCs are nonzero? In scenario (iv), four WCs --
$C_{9,\NP}^{\mu\mu}$, $C_{10,\NP}^{\mu\mu}$
$C_{9,\NP}^{\prime\mu\mu}$, and $C_{10,\NP}^{\prime\mu\mu}$ -- are
allowed to be nonzero. We find the best-fit values of the WCs and the
prediction for $R_{K^*}^{low}$ to be
\bea
{\rm (iv)} && C_{9,\NP}^{\mu\mu} = -1.10 \pm 0.22 ~,~~ C_{10,\NP}^{\mu\mu} = 0.28 \pm 0.17 ~, \\
&& C_{9,\NP}^{\prime\mu\mu} = 0.11 \pm 0.45 ~,~~ C_{10,\NP}^{\prime\mu\mu} = -0.21 \pm 0.30 : 
R_{K^*}^{low} = (0.83)~0.85 ~. \nn
\label{scenario(iv)}
\eea
Here the smallest predicted value of $R_{K^*}^{low}$ (the number in
parentheses) is computed as follows. In scenarios (i)-(iii), we have
determined that varying $C_{9,\NP}^{\mu\mu}$ and $C_{10,\NP}^{\mu\mu}$
does not significantly affect $R_{K^*}^{low}$. Thus, for simplicity,
we set these WCs equal to their best-fit values. The smallest
predicted value of $R_{K^*}^{low}$ is then found by scanning the 68\%
C.L.\ region in $C_{9,\NP}^{\prime\mu\mu}$-$C_{10,\NP}^{\prime\mu\mu}$
space. But even in this case, the predicted value of $R_{K^*}^{low}$
is still quite a bit larger than the measured value. This leads us to
conclude that if there is NP only in $\bsmumu$, $R_{K^*}^{low} \ge
0.83$ is predicted, which is more than $1.5\sigma$ above its measured
value\footnote{We note that, if all four WCs ($C_{9,10,\NP}^{\mu\mu}$,
  $C_{9,10,\NP}^{\prime\mu\mu}$) are allowed to vary, one can generate
  a smaller value of $R_{K^*}^{low}$, 0.81. This is due only to the
  fact that the allowed region in the space of WCs is considerably
  larger: when one varies two parameters, the 68\% C.L.\ region is
  defined by $\chi^2 \le \chi^2_{\rm min} + 2.3$, whereas when one
  varies four parameters, it is $\chi^2 \le \chi^2_{\rm min} +
  4.72$.}.

Of course, when one tries to simultaneously explain a number of
different observables, it is not necessary that every experimental
result be reproduced within $1\sigma$. As long as the overall fit has
$\chi^2_{\rm min}/d.o.f.\ \sim 1$, it is considered acceptable. This
is indeed what is found in the analyses in which NP is assumed to be
only in $\bsmumu$ \cite{Capdevila:2017bsm, Altmannshofer:2017yso,
  DAmico:2017mtc, Hiller:2017bzc, Geng:2017svp, Ciuchini:2017mik,
  Celis:2017doq, Alok:2017sui}. Still, this raises the question:
suppose that the true value of $R_{K^*}^{low}$ is near its
measured value. What is required to explain it?

This has been explored in a few papers. In Refs.~\cite{Datta:2017ezo,
  Altmannshofer:2017bsz}, it is argued that $R_{K^*}^{low}$ cannot be
explained by new short-distance interactions, so that a very light
mediator is required, with a mass in the 1-100 MeV range. And in
Ref.~\cite{Bardhan:2017xcc}, it is said that $R_{K^*}^{low}$ cannot be
reproduced with only vector and axial vector operators, leading to the
suggestion of tensor operators. In the present paper, we show that, in
fact, one {\it can} generate a value for $R_{K^*}^{low}$ near its
measured value with short-range interactions involving vector and
axial vector operators.

To be specific, we show that, if there are NP contributions to
$\bsee$, one can account for $R_{K^*}^{low}$.\footnote{NP in $\bsee$
  has also been considered in some previous studies. In
  Refs.~\cite{Capdevila:2017bsm, Altmannshofer:2017yso, Geng:2017svp},
  it is found that the $R_{K^{(*)}}$ data can be explained by NP in
  $\bsmumu$ or $\bsee$. A more complete analysis, similar to that
  performed in the present paper, is carried out in
  Ref.~\cite{Ciuchini:2017mik}. However, there they do not focus on
  $R_{K^*}^{low}$.} Using a model-independent, effective-field-theory
approach, we find that there are quite a few scenarios involving
various NP WCs in $\bsmumu$ and $\bsee$ in which a value for
$R_{K^*}^{low}$ can be generated that is larger than its measured
value, but within $\sim 1\sigma$.  Indeed, if there is NP in
$\bsmumu$, it is not a stretch to imagine that it also contributes to
$\bsee$. We consider the most common types of NP models that have been
proposed to explain the $\bsmumu$ anomalies -- those containing
leptoquarks or a $Z'$ gauge boson -- and find that, if they are
allowed to contribute to $\bsee$, the measured value of
$R_{K^*}^{low}$ can be accounted for (within $\sim 1\sigma$).

In scenario (ii) above, $C_{9,\NP}^{\mu\mu} = -C_{10,\NP}^{\mu\mu}$,
so the NP couples only to the left-handed (LH) quarks and $\mu$. This
is a popular scenario, and many models have been constructed that have
purely LH couplings. However, we find that, if the NP couplings in
$\bsee$ are also purely LH, $R_{K^*}^{low}$ can {\it not} be explained
-- couplings involving the right-handed (RH) quarks and/or leptons
must be involved.

One feature of this type of NP is that it is independent of $q^2$.
Thus, if the $\bsee$ WCs are affected in a way that lowers the value
of $R_{K^*}^{low}$ compared to what is found if the NP affects only
$\bsmumu$, the value of $R_{K^*}^{cen}$ is also lowered. We generally
find that, if the true value of $R_{K^*}^{low}$ is $\sim 1\sigma$
above its present measured value, the true value of $R_{K^*}^{cen}$
will be found to be $\sim 1\sigma$ below its present measured value.
This is a prediction of this NP explanation.

As noted above, there are a number of scenarios involving different
sets of $\bsmumu$ and $\bsee$ NP WCs in which $R_{K^*}^{low}$ can be
explained. Since NP in $\bsee$ is independent of $q^2$, each of these
scenarios makes specific predictions for the values of $R_{K^*}$ and
$R_K$ in other $q^2$ bins. Furthermore, a future precise measurement
of the LFUV observable $Q_5 \equiv P^{\prime\mu\mu}_5 - P^{\prime
  ee}_5$ will help to distinguish the various scenarios.

The observables in $B \to K^* \mu^+\mu^-$ and $B_s^0 \to \phi \mu^+
\mu^-$ are Lepton-Flavour Dependent (LFD), while $R_K$ and $R_{K^*}$
are Lepton-Flavour-Universality-Violating (LFUV) observables. If one
assumes NP only in $\bsmumu$, one uses LFUV NP to explain both LFD and
LFUV observables. Recently, in Ref.~\cite{Alguero:2018nvb},
Lepton-Flavour-Universal (LFU) NP was added. The LFUV observables are
then explained by the LFUV NP, while the LFD observables are explained
by LFUV $+$ LFU NP. Our scenarios, with NP in $\bsmumu$ and $\bsee$,
can be translated into LFUV $+$ LFU NP, and vice-versa. As we will
see, the two ways of categorizing the NP are complementary to one
another.

We begin in Sec.~2 with a detailed discussion of how the addition of
NP in $\bsee$ can explain $R_{K^*}^{low}$. We construct a number of
different scenarios using both a model-independent,
effective-field-theory approach, and within specific models involving
leptoquarks or a $Z'$ gauge boson. In Sec.~3, we examine the
predictions of the various scenarios for $R_{K^{(*)}}$ and $Q_5$, and
compare NP in $\bsmumu$ and $\bsee$ to LFUV $+$ LFU NP. We conclude in
Sec.~4.

\section{\boldmath NP in $\bsmumu$ and $\bsee$}

We repeat the fit, but allowing for NP in both $\bsmumu$ and $\bsee$
transitions. The $\bsmumu$ observables used in the fit are given in
Ref.~\cite{Alok:2017sui}. The $\bsee$ observables that have been
measured are given in Table \ref{NP_RK*lowq2} \cite{futurebsee}. In
this Table, we see that most observables have sizeable errors.  The
one exception is $\frac{d \cB}{dq^2}(B^+ \to K^+ e^+ e^-)$, but here
the theoretical uncertainties are significant. The net effect is that
NP in $\bsee$ is rather less constrained than NP in $\bsmumu$.

Note that $P_4^{\prime e}$ and $P_5^{\prime e}$ have been measured in
two different ranges of $q^2$, [0.1-4.0] GeV$^2$ and [1.0-6.0]
GeV$^2$. These regions overlap, so including both measurements in the
fit would be double counting. Since we are interested in the
predictions for $R_{K^*}^{low}$, in the fit we use the observables for
$q^2$ in the lower range, [0.1-4.0] GeV$^2$. However, we have verified
that the results are little changed if we use the observables for
$q^2$ in the other range, [1.0-6.0] GeV$^2$.

\begin{table*}[t] 
\renewcommand{\arraystretch}{2}
\begin{center}
\begin{tabular}{|c|c|c|}
\hline
Observables & $q^2$ $\rm (GeV^2)$ & Measurement\\ \hline 
$P_4^{\prime e}$  &[0.1-4.0] & $0.34^{+0.41}_{-0.45} \pm 0.11$  \cite{Wehle:2016yoi}\\ 
$P_5^{\prime e}$   & [0.1-4.0]&  $0.51^{+0.39}_{-0.46}\pm 0.09$  \cite{Wehle:2016yoi}\\ 
$P_4^{\prime e}$  &[1.0-6.0] & $-0.72^{+0.40}_{-0.39} \pm 0.06$  \cite{Wehle:2016yoi}\\ 
$P_5^{\prime e}$   & [1.0-6.0]&  $-0.22^{+0.39}_{-0.41}\pm 0.03$  \cite{Wehle:2016yoi}\\ 
$P_4^{\prime e}$  &[14.18-19.0] & $-0.15^{+0.41}_{-0.40} \pm 0.04$  \cite{Wehle:2016yoi}\\ 
$P_5^{\prime e}$  &[14.18-19.0] & $-0.91^{+0.36}_{-0.30}\pm0.03$  \cite{Wehle:2016yoi}\\
$\frac{d \cB}{dq^2}(B^0\to K^* e^+ e^-)$  &[0.001-1.0] & $(3.1^{+0.9}_{-0.8} \pm 0.2) \times 10^{-7}$ \cite{Aaij:2013hha}\\  
$F_L(B^0\to K^* e^+ e^-)$  &[0.002-1.12] &  $0.16\pm 0.06 \pm 0.03$   \cite{Aaij:2015dea}\\
$\cB(B \to X_s e^+ e^-)$  &[1.0-6.0] & $(1.93^{+0.47+0.21}_{-0.45-0.16}\pm 0.18)\times 10^{-6}$ \cite{Lees:2013nxa}\\ 
$\cB(B \to X_s e^+ e^-)$  &[14.2-25.0] & $(0.56^{+0.19+0.03}_{-0.18-0.03})\times 10^{-6}$ \cite{Lees:2013nxa}\\
$\frac{d \cB}{dq^2}(B^+ \to K^+ e^+ e^-)$  &[1.0-6.0] & $(0.312^{ +0.038+0.012}_{-0.030-0.008} )\times 10^{-7}$ \cite{RKexpt}\\
\hline
\end{tabular}
\end{center}
\caption{Measured $\bsee$ observables.}
\label{NP_RK*lowq2}
\end{table*}

The fit can be done in two different ways. First, there is the
model-independent, effective-field-theory approach. Here, the NP WCs
are all taken to be independent. The fit is performed simply assuming
that certain WCs in $\bsmumu$ and $\bsee$ transitions are nonzero,
without addressing what the underlying NP model might be. Second, in
the model-dependent approach, the fit is performed in the context of a
specific model. Since the NP WCs are all functions of the model
parameters, there may be relations among the WCs, i.e., they may not
all be independent. Furthermore, there may be additional constraints
on the model parameters due to other processes. Each approach has
certain advantages, and, in the subsections below, we consider both of
them.

\subsection{Model-independent Analysis}

In this subsection, we examine several different cases with $m + n$ NP
WCs, where $m$ and $n$ are respectively the number of independent NP
WCs (or combinations of WCs) in $\bsmumu$ and $\bsee$. For each case,
we find the best-fit values of the NP WCs, and compute the prediction
for $R_{K^*}^{low}$. 

\subsubsection{Cases with $1+1$ NP WCs}

Here we consider the simplest case, in which there is one nonzero NP
WC (or combination of WCs) in each of $\bsmumu$ and $\bsee$. We are
looking for scenarios that satisfy the following condition: if one
varies the NP WCs within their 68\% C.L.-allowed region (taking into
accout the fact that the errors on the WCs are correlated), one can
generate a value for $R_{K^*}^{low}$ that is within $\sim 1\sigma$ of
its measured value.

Although many of the scenarios we examined do not satisfy this
conditon, we found several that do. They are presented in the first
four entries of Table \ref{NP_RK*lowq2_1}. In each scenario, the
right-hand number in the $R_{K^*}^{low}$ column is its predicted value
for the best-fit value of the WCs. The number in parentheses to the
left is the smallest predicted value of $R_{K^*}^{low}$ within the
$1\sigma$ (68\% C.L.)  range of the WCs. The $R_{K^*}^{cen}$ and $R_K$
columns are similar, except that the numbers in parentheses are the
values of $R_{K^*}^{cen}$ and $R_K$ evaluated at the point that yields
the smallest value of $R_{K^*}^{low}$. We also examine how much better
than the SM each scenario is at explaining the data. This is done by
computing the pull $= \sqrt{\chi^2_{SM} - \chi^2_{SM + NP}}$,
evaluated using the best-fit values of the WCs.

\begin{table*}[t] 
\renewcommand{\arraystretch}{2}
\begin{center}
\begin{tabular}{|c|l|l|c|c|c|c|}
 \hline 
& NP in $\bsmumu$ & NP in $\bsee$ & $R_{K^*}^{low}$ & $R_{K^*}^{cen}$ & $R_K$ & Pull \\
\hline
S1 & $C_{9,\NP}^{\mu\mu} = -C_{10,\NP}^{\mu\mu}$ & $C_{10,\NP}^{ee} = -C_{10,\NP}^{'ee}$ & & & & \\
& ~~~$= -0.57 \pm 0.09$ 
& ~~~$= -0.25 \pm 0.27$ & 
(0.76) 0.82 &
(0.54) 0.66 &
(0.76) 0.74 &
6.5 \\
\hline
S2 & $C_{9,\NP}^{\mu\mu} = -C_{9,\NP}^{'\mu\mu}$ & $C_{9,\NP}^{'ee} = C_{10,\NP}^{'ee}$ & & & & \\
& ~~~$= -0.95 \pm 0.17$ 
& ~~~$= -1.7 \pm 0.30$ &
(0.75) 0.82 &
(0.52) 0.65 &
(0.77) 0.82 &
6.5 \\
\hline
S3 & $C_{9,\NP}^{\mu\mu}$ & $C_{9,\NP}^{ee} = -C_{9,\NP}^{'ee}$ & & & & \\ 
& ~~~$= -1.10 \pm 0.17$ & ~~~$= 0.52 \pm 0.31$ &
(0.78) 0.83 &
(0.58) 0.68 &
(0.77) 0.77 &
6.6 \\
\hline
S4 & $C_{9,\NP}^{\mu\mu}$ & $C_{10,\NP}^{ee} = -C_{10,\NP}^{'ee}$ & & & & \\
& ~~~$= -1.06 \pm 0.17$ & ~~~$= -0.44 \pm 0.26$ &
(0.78) 0.82 &
(0.58) 0.67 &
(0.77) 0.78 &
6.7 \\
\hline
\hline
S5 & $C_{9,\NP}^{\mu\mu} = -C_{10,\NP}^{\mu\mu}$ & $C_{9,\NP}^{ee} = C_{10,\NP}^{ee}$ & & & & \\
& ~~~$= -0.51 \pm 0.12$ 
& ~~~$= -0.66 \pm 0.55$ & 
(0.80) 0.83 &
(0.64) 0.70 &
(0.70) 0.74 &
6.4 \\
\hline
S6 & $C_{9,\NP}^{\mu\mu} = -C_{10,\NP}^{\mu\mu}$ & $C_{9,\NP}^{\prime ee} = C_{10,\NP}^{\prime ee}$ & & & & \\
& ~~~$= -0.64 \pm 0.10$ 
& ~~~$= 0.42 \pm 0.89$ & 
(0.81) 0.85 &
(0.64) 0.70 &
(0.68) 0.71 &
6.3 \\
\hline
\hline
S7 & $C_{9,\NP}^{\mu\mu} = -C_{10,\NP}^{\mu\mu}$ & $C_{9,\NP}^{ee} = -C_{10,\NP}^{ee}$ & & & & \\
& ~~~$= -0.65 \pm 0.12$ 
& ~~~$= -0.06 \pm 0.18$ & 
(0.85) 0.86 &
(0.73) 0.74 &
(0.73) 0.73 &
6.4 \\
\hline
\end{tabular}
\end{center}
\caption{Scenarios with one nonzero NP WC (or combination of WCs) in
  each of $\bsmumu$ and $\bsee$, and their predictions for
  $R_{K^*}^{low}$, $R_{K^*}^{cen}$ and $R_K$. The pulls for each
  scenario are also shown.}
\label{NP_RK*lowq2_1}
\end{table*}

In all four scenarios, the addition of NP in $\bsee$ makes it possible
to produce a value of $R_{K^*}^{low}$ roughly $1\sigma$ above its
measured value, which is an improvement on the situation where the NP
affects only $\bsmumu$. As noted in the introduction, this type of NP
is independent of $q^2$, so that, if one adds NP to $\bsee$ in a way
that lowers the predicted value of $R_{K^*}^{low}$, it will also lower
the predicted value of $R_{K^*}^{cen}$. Indeed, we see that the values
of the NP WCs that produce a better value of $R_{K^*}^{low}$ also lead
to a value of $R_{K^*}^{cen}$ that is roughly $1\sigma$ below its
measured value. This is then a prediction: if the true value of
$R_{K^*}^{low}$ is near its measured value, and if this is due to NP
in $\bsee$, the true value of $R_{K^*}^{cen}$ will be found to be
below its measured value.

Note that this behaviour does not apply to $R_K$. Its measured value
is \cite{RKexpt}
\beq
R_K^\expt = 0.745^{+0.090}_{-0.074}~{\rm (stat)} \pm 0.036~{\rm (syst)} ~,
\eeq
which differs from the SM prediction of $R_K^\SM = 1 \pm 0.01$
\cite{IsidoriRK} by $2.6\sigma$. In all scenarios, the value of
$R_K^\expt$ is accounted for, and this changes little if one uses the
central values of the NP WCs or the values that lead to a lower
$R_{K^*}^{low}$.

The pulls for all four scenarios are sizeable and roughly equal. It
must be stressed that the values of pulls are strongly dependent on
how the analysis is done: what observables are included, how
theoretical errors are treated, which form factors are used, etc. For
this reason one must be very careful in comparing pulls found in
different analyses. On the other hand, comparing the pulls of various
scenarios within a single analysis may be illuminating. With this in
mind, consider again scenarios (i) and (ii) [Eq.~(\ref{bsmumuNP})],
and compare them with scenarios S3 and S1, respectively, of Table
\ref{NP_RK*lowq2_1}. Below we present the pulls of (i) and
(ii)\footnote{In Ref.~\cite{Altmannshofer:2017fio}, using only
  $\bsmumu$ data (i.e., $R_{K^{(*)}}$ data was not included), the
  pulls of (i) and (ii) were found to be 5.2 and 4.8, respectively.
  Using the same method of analysis, we added the $R_{K^{(*)}}$ data
  and found that the pulls were increased to 6.2 and 6.3,
  respectively.}, and repeat some information given previously, in
order to facilitate the comparison:
\bea
{\rm (i)} ~~ C_{9,\NP}^{\mu\mu} = -1.20 &~:~&
~~ R_{K^*}^{low} = 0.89 ~,~~ R_{K^*}^{cen} = 0.83 ~,~~ R_K = 0.76 ~,~~ {\rm pull} = 6.2 ~, \nn\\
S3 ~~ C_{9,\NP}^{\mu\mu} = -1.10 &~:~& 
~~ R_{K^*}^{low} = 0.83 ~,~~ R_{K^*}^{cen} = 0.68 ~,~~ R_K = 0.77 ~,~~ {\rm pull} = 6.6 ~, \nn\\
{\rm (ii)} ~~ C_{9,\NP}^{\mu\mu} = -C_{10,\NP}^{\mu\mu} = -0.62 &~:~&
~~ R_{K^*}^{low} = 0.85 ~,~~ R_{K^*}^{cen} = 0.73 ~,~~ R_K = 0.72 ~,~~ {\rm pull} = 6.3 ~, \nn\\
S1 ~~ C_{9,\NP}^{\mu\mu} = -C_{10,\NP}^{\mu\mu} = -0.57 &~:~&
~~ R_{K^*}^{low} = 0.82 ~,~~ R_{K^*}^{cen} = 0.66 ~,~~ R_K = 0.74 ~,~~ {\rm pull} = 6.5 ~, \nn\\
{\rm experiment} &~:~& ~~ R_{K^*}^{low} = 0.66 ~,~~ R_{K^*}^{cen} = 0.69 ~,~~ R_K = 0.75 ~.
\eea
We first compare scenarios (i) and S3, noting that pull[S3] $>$
pull[(i)]. What is this due to? In the two scenarios, the value of
$C_{9,\NP}^{\mu\mu}$ is very similar, so that the contribution to the
pull of the $\bsmumu$ observables is about the same in both cases.
(Indeed, the dominant source of the large pull is NP in $\bsmumu$.)
That is, the difference in the pulls is due to the addition of NP in
$\bsee$ in S3. Now, the $\bsee$ observablies in Table
\ref{NP_RK*lowq2} have virtually no effect on the pull; the important
effect is the different predictions for $R_{K^{(*)}}$. Above, we see
that the prediction of scenario S3 for $R_{K^*}^{cen}$
($R_{K^*}^{low}$) is much (slightly) closer to the experimental value
than that of scenario (i). (The predictions for $R_K$ are essentially
the same.) This leads to an increase of 0.4 in the pull. The
comparison of scenarios (ii) and S1 is similar.

We also note that, in all scenarios, the pull of the fits evaluated at
the (68\% C.L.) point that yields the smallest value of
$R_{K^*}^{low}$ is only $\sim 0.2$ smaller than the central-value
pull.  That is, if NP is added to the $\bsee$ WCs, it costs very
little in terms of the pull to improve the agreement with the measured
value of $R_{K^*}^{low}$.

In scenario S5 of Table \ref{NP_RK*lowq2_1}, when the NP is integrated
out, the four-fermion operators $[ {\bar s} \gamma_\mu P_L b ] [
  {\bar\mu} \gamma^\mu P_L \mu ]$ and $[ {\bar s} \gamma_\mu P_L b ] [
  {\bar e} \gamma^\mu P_R e ]$ are generated. That is, the NP couples
to the LH quarks and $\mu$, but to the RH $e$. In scenario S6, one has
the four-fermion operators $[ {\bar s} \gamma_\mu P_L b ] [ {\bar\mu}
  \gamma^\mu P_L \mu ]$ and $[ {\bar s} \gamma_\mu P_R b ] [ {\bar e}
  \gamma^\mu P_R e ]$, so that the NP couples to the LH quarks and
$\mu$, but to the RH quarks and $e$. We have not included either of
these among the satisfactory scenarios, since the smallest value of
$R_{K^*}^{low}$ possible at 68\% C.L.\ is 0.80 or 0.81, which are a
bit larger than $1\sigma$ above the measured value of
$R_{K^*}^{low}$. However, it must be conceded that this cutoff is
somewhat arbitrary, so that these scenarios, and others like them,
should be considered borderline.

Finally, in scenario S7 of Table \ref{NP_RK*lowq2_1}, the NP
four-fermion operators are $[ {\bar s} \gamma_\mu P_L b ] [ {\bar\mu}
  \gamma^\mu P_L \mu ]$ and $[ {\bar s} \gamma_\mu P_L b ] [ {\bar e}
  \gamma^\mu P_L e ]$, i.e., the NP couples only to LH particles. This
is a popular choice for model builders. However, here the smallest
predicted value for $R_{K^*}^{low}$ is still almost $2\sigma$ above
its measured value, so this cannot be considered a viable scenario.

\subsubsection{Cases with more than $1+1$ NP WCs}

We now consider more general scenarios, in which there are $m$ ($n$)
nonzero NP WCs (or combinations of WCs) in $\bsmumu$ ($\bsee$), with
$m \ge 1$, $n \ge 1$ and $m + n > 2$. As discussed in the
introduction, we know that varying the $\bsmumu$ NP WCs has little
effect on $R_{K^*}^{low}$. We therefore fix these WCs to their central
values and vary the $\bsee$ NP WCs within their 68\% C.L.-allowed
region to obtain the smallest predicted value of $R_{K^*}^{low}$. We
find that there are now many solutions that predict a value for
$R_{K^*}^{low}$ that is within roughly $1\sigma$ of its measured
value. In Table \ref{NP_RK*lowq2_2} we present four of these.
Scenarios S8 and S9 have $m = 1$ and $n = 2$, while scenarios S10
and S11 have $m = n = 2$. 

\begin{table*}[t] 
\renewcommand{\arraystretch}{2}
\begin{center}
\begin{tabular}{|c|l|l|c|c|c|c|}
 \hline 
& NP in $\bsmumu$ & NP in $\bsee$ & $R_{K^*}^{low}$ & $R_{K^*}^{cen}$ & $R_K$ & Pull \\
\hline
S8 & $C_{9,\NP}^{\mu\mu} = -C_{10,\NP}^{\mu\mu}$ & $C_{9,\NP}^{ee} = -1.0 \pm 1.0$ & & & & \\
& ~~~~~~~ $= -0.52 \pm 0.14$ & 
$C_{10,\NP}^{ee} = -0.81 \pm 0.58$ &
(0.79) 0.83 &
(0.61) 0.69 &
(0.69) 0.75 & 6.5 \\
\hline
S9 & $C_{9,\NP}^{\mu\mu} = -C_{10,\NP}^{\mu\mu}$ & $C_{9,\NP}^{'ee} = 1.00 \pm 0.65$ & & & & \\
& ~~~~~~~ $= -0.52 \pm 0.12$ & 
$C_{10,\NP}^{'ee} = 1.24 \pm 0.76$ &
(0.75) 0.82 &
(0.53) 0.65 &
(0.79) 0.76 & 6.4 \\
\hline
\hline
S10 & $C_{9,\NP}^{\mu\mu} = -0.96 \pm 0.22$ & $C_{9,\NP}^{ee} = -1.23 \pm 1.01$ & & & & \\
& $C_{10,\NP}^{\mu\mu} = 0.24 \pm 0.22$ & $C_{10,\NP}^{ee} = -0.84 \pm 0.53$ & 
(0.78) 0.84 &
(0.59) 0.71 &
(0.63) 0.75 & 6.8 \\
\hline
S11 & $C_{9,\NP}^{\mu\mu} = -1.08 \pm 0.22$ & $C_{9,\NP}^{'ee} = 0.67 \pm 0.91$ & & & & \\ 
& $C_{10,\NP}^{\mu\mu} = 0.26 \pm 0.22$ & $C_{10,\NP}^{'ee} = 1.04 \pm 0.99$ &
(0.77) 0.83 &
(0.55) 0.66 &
(0.77) 0.76 & 6.8 \\
\hline
\end{tabular}
\end{center}
\caption{Scenarios with $m$ ($n$) nonzero NP WCs (or combinations of
  WCs) in $\bsmumu$ ($\bsee$), with $m \ge 1$, $n \ge 1$ and $m + n >
  2$, that can generate a value for $R_{K^*}^{low}$ within $\sim
  1\sigma$ of its measured value. Predictions for $R_{K^*}^{cen}$ and
  $R_K$, as well as the pulls for each scenario, are also shown.}
\label{NP_RK*lowq2_2}
\end{table*}

We see that, despite having a larger number of nonzero independent NP
WCs, at 68\% C.L.\ these scenarios predict similar values for
$R_{K^*}^{low}$ as the scenarios in Table \ref{NP_RK*lowq2_1}.
Furthermore, the NP WCs that produce these values for $R_{K^*}^{low}$
also predict values for $R_{K^*}^{cen}$ that are below its measured
value. Finally, as was the case for scenarios with $1+1$ NP WCs, all
scenarios here explain $R_K^\expt$, even for values of the NP WCs that
lead to a lower $R_{K^*}^{low}$.

As was the case with the scenarios of Table \ref{NP_RK*lowq2_1}, here
the pulls are again sizeable. And again, it is interesting to compare
similar scenarios without and with NP in $\bsee$. Consider scenarios
(iii) [Eq.~(\ref{scenario(iii)})] and S10:
\bea
{\rm (iii)} ~~ C_{9,\NP}^{\mu\mu} = -1.10 ~,~~ C_{10,\NP}^{\mu\mu} = 0.28 &~:~&
R_{K^*}^{low} = 0.87 ~,~~ R_{K^*}^{cen} = 0.74 ~,~~ R_K = 0.71 ~,~~ {\rm pull} = 6.6 ~, \nn\\
S10 ~~ C_{9,\NP}^{\mu\mu} = -0.96 ~,~~ C_{10,\NP}^{\mu\mu} = 0.24 &~:~& 
R_{K^*}^{low} = 0.84 ~,~~ R_{K^*}^{cen} = 0.71 ~,~~ R_K = 0.75 ~,~~ {\rm pull} = 6.8 ~, \nn\\
{\rm experiment} &~:~& R_{K^*}^{low} = 0.66 ~,~~ R_{K^*}^{cen} = 0.69 ~,~~ R_K = 0.75 ~.
\eea
The values of the $\bsmumu$ NP WCs are very similar in the two
scenarios, so that the difference in pulls is due principally to the
addition of NP in $\bsee$ in S10. Looking at $R_{K^{(*)}}$, we see
that the predictions of scenario S10 for $R_{K^*}^{low}$,
$R_{K^*}^{cen}$ and $R_K$ are all slightly closer to the experimental
values than the predictions of (iii). This leads to an increase of 0.2
in the pull.

\subsection{Model-dependent Analysis}

There are two types of NP models in which there is a tree-level
contribution to $\bsmumu$: those containing leptoquarks (LQs), and
those with a $Z'$ boson. In this subsection, we examine these models
with the idea of explaining $R_{K^*}^{low}$ by adding a contribution
to $\bsee$. To be specific, we want to answer the question: can the
scenarios in Tables \ref{NP_RK*lowq2_1} and \ref{NP_RK*lowq2_2} be
reproduced within LQ or $Z'$ models? In the following, we examine
these two types of NP models.

\subsubsection{Leptoquarks}
\label{LQsection}

There are ten LQ models that couple to SM particles through dimension
$\le 4$ operators \cite{AGC}. There include five spin-0 and five
spin-1 LQs, denoted $\Delta$ and $V$ respectively. Their couplings are
\bea
{\cal L}_\Delta & = & ( y_{\ell u} {\bar\ell}_L u_R + y_{eq}\, {\bar e}_R i \tau_2 q_L ) \Delta_{-7/6}
+ y_{\ell d}\, {\bar\ell}_L d_R \Delta_{-1/6}
+ ( y_{\ell q}\, {\bar\ell}^c_L i \tau_2 q_L + y_{eu} \, {\bar e}^c_R u_R ) \Delta_{1/3} \nn\\
&& +~y_{ed}\, {\bar e}^c_R d_R \Delta_{4/3}
+ y'_{\ell q}\, {\bar\ell}^c_L i \tau_2 {\vec \tau} q_L \cdot {\vec \Delta}'_{1/3} + h.c. \nn\\
{\cal L}_V & = & (g_{\ell q}\, {\bar\ell}_L \gamma_\mu q_L + g_{ed}\, {\bar e}_R \gamma_\mu d_R) V^\mu_{-2/3}
+ g_{eu} \, {\bar e}_R \gamma_\mu u_R V^\mu_{-5/3}
+ g'_{\ell q}\, {\bar\ell}_L \gamma_\mu {\vec \tau} q_L \cdot {\vec V}^{\prime \mu}_{-2/3} \nn\\
&& +~(g_{\ell d}\, {\bar\ell}_L \gamma_\mu d_R^c + g_{eq}\, {\bar e}_R \gamma_\mu q^c_L) V^\mu_{-5/6}
+ + g_{\ell u} \, {\bar \ell}_L \gamma_\mu u_R^c V^\mu_{1/6} + h.c.,
\label{LQlist}
\eea
where, in the fermion currents and in the subscripts of the couplings,
$q$ and $\ell$ represent left-handed quark and lepton $SU(2)_L$
doublets, respectively, while $u$, $d$ and $e$ represent right-handed
up-type quark, down-type quark and charged lepton $SU(2)_L$ singlets,
respectively. The subscripts of the LQs indicate the hypercharge,
defined as $Y = Q_{em} - I_3$.

In the above, the LQs can couple to fermions of any generation. To
specify which particular fermions are involved, we add superscripts to
the couplings. For example, $g^{\prime \mu s}_{\ell q}$ is the
coupling of the ${\vec V}^{\prime \mu}_{-2/3}$ LQ to a left-handed
$\mu$ (or $\nu_\mu$) and a left-handed $s$ (or $c$). Similarly, $y_{e
  q}^{e b}$ is the coupling of the $\Delta_{-7/6}$ LQ to a
right-handed $e$ and a left-handed $b$. These couplings are relevant
for $\bsmumu$ or $\bsee$ (and possibly $\bsnunubar$). Note that the
$\Delta_{1/3}$, $V^\mu_{-5/3}$ and $V^\mu_{1/6}$ LQs do not contribute
to $\bsll$. In Ref.~\cite{Sakakietal}, ${\vec \Delta}'_{1/3}$,
$V^\mu_{-2/3}$ and ${\vec V}^{\prime \mu}_{-2/3}$ are called $S_3$,
$U_1$ and $U_3$, respectively, and we adopt this nomenclature below.

In a model-dependent analysis, one must take into account the fact
that, within a particular model, there may be contributions to
additional observables. In the case of LQ models, in addition to
$O^{(\prime)\ell\ell}_{9,10}$ ($\ell = e,\mu$) [Eq.~(\ref{Heff})],
there may be contributions to the lepton-flavour-conserving operators
\bea
& O^{(\prime)\ell\ell}_\nu = [ {\bar s} \gamma_\mu P_{L(R)} b ] [ {\bar\nu}_\ell \gamma^\mu (1 - \gamma_5) \nu_\ell ] ~, & \nn\\
& O^{(\prime)\ell\ell}_S = [ {\bar s} P_{R(L)} b ] [ {\bar\ell} \ell ] ~~,~~~~
O^{(\prime)\ell\ell}_P = [ {\bar s} P_{R(L)} b ] [ {\bar\ell} \gamma_5 \ell ] & ~.
\label{newops}
\eea
$O^{(\prime)\ell\ell}_\nu$ contributes to $b \to s \nu_\ell {\bar
  \nu}_\ell$, while $O^{(\prime)\ell\ell}_S$ and
$O^{(\prime)\ell\ell}_P$ are additional contributions to $b \to s
\ell^+ \ell^-$. There may also be contributions to the
lepton-flavour-violating (LFV) operators
\bea
& O^{(\prime)\ell\ell'}_{9(10)} = [ {\bar s} \gamma_\mu P_{L(R)} b ] [ {\bar\ell} \gamma^\mu (\gamma_5) \ell' ] ~, & \nn\\
& O^{(\prime)\ell\ell'}_\nu = [ {\bar s} \gamma_\mu P_{L(R)} b ] [ {\bar\nu}_\ell \gamma^\mu (1 - \gamma_5) \nu_{\ell'} ] ~, & \nn\\
& O^{(\prime)\ell\ell'}_S = [ {\bar s} P_{R(L)} b ] [ {\bar\ell} \ell' ] ~~,~~~~
O^{(\prime)\ell\ell'}_P = [ {\bar s} P_{R(L)} b ] [ {\bar\ell} \gamma_5 \ell' ] & ~,
\label{newopsLFV}
\eea
where $\ell,\ell' = e,\mu$, with $\ell \ne \ell'$.
$O^{(\prime)\ell\ell'}_{9(10)}$, $O^{(\prime)\ell\ell'}_S$ and
$O^{(\prime)\ell\ell'}_P$ contribute to $\bs \to e^\pm \mu^\mp$ and $B
\to K^{(*)} e^\pm \mu^\mp$. Using the couplings in Eq.~(\ref{LQlist}),
one can compute which WCs are affected by each LQ. These are shown in
Table~\ref{LQWC} for $\ell = \ell' = \mu$ \cite{AGC}, and it is
straightforward to change one $\mu$ or both to an $e$. Finally, there
may also be a 1-loop contribution to the LFV decay $\mu \to e \gamma$:
\beq
O^{(L)R}_\gamma = [ {\bar e} \sigma_{\mu \nu} P_{L(R)} \mu ] F_{\mu\nu} ~.
\label{ogamma}
\eeq
All LFV operators can arise if there is a single LQ that couples to
both $\mu$ and $e$. However, if two different LQs couple to $\mu$ and
$e$, there are no contributions to LFV processes. Since the
constraints from LFV processes are extremely stringent, we therefore
anticipate that it will be difficult to explain $R_{K^*}^{low}$ in a
model with a single LQ.

\begin{table}
\begin{center}
\begin{tabular}{|c|cccc|} \hline
LQ & $C_{9,\NP}^{\mu\mu}$ & $C_{10,\NP}^{\mu\mu}$ & $C_{9,\NP}^{\prime\mu\mu}$ & $C_{10,\NP}^{\prime\mu\mu}$ \\
   & $C_{S,\NP}^{\mu\mu}$ & $C_{S,\NP}^{\prime\mu\mu}$ & $C_{\nu,\NP}^{\mu\mu}$ & $C_{\nu,\NP}^{\prime\mu\mu}$ \\
\hline
${\vec \Delta}'_{1/3} ~[S_3]$ & $y_{\ell q}^{\prime \mu b} (y_{\ell q}^{\prime \mu s})^*$
                             & $- y_{\ell q}^{\prime \mu b} (y_{\ell q}^{\prime \mu s})^*$ & 0 & 0 \\
& 0 & 0 & $\frac12 y_{\ell q}^{\prime \mu b} (y_{\ell q}^{\prime \mu s})^*$ & 0 \\
\hline
$\Delta_{-7/6}$ & $-\frac12 y_{e q}^{\mu b} (y_{e q}^{\mu s})^*$ & $-\frac12 y_{e q}^{\mu b} (y_{e q}^{\mu s})^*$ & 0 & 0 \\
& 0 & 0 & 0 & 0 \\
\hline
$\Delta_{-1/6}$ & 0 & 0 & $-\frac12 y_{\ell d}^{\mu b} (y_{\ell d}^{\mu s})^*$ & $\frac12 y_{\ell d}^{\mu b} (y_{\ell d}^{\mu s})^*$ \\
& 0 & 0 & 0 & $-\frac12 y_{\ell d}^{\mu b} (y_{\ell d}^{\mu s})^*$ \\
\hline
$\Delta_{4/3}$ & 0 & 0 & $\frac12 y_{e d}^{\mu b} (y_{e d}^{\mu s})^*$ & $\frac12 y_{e d}^{\mu b} (y_{e d}^{\mu s})^*$ \\
& 0 & 0 & 0 & 0 \\
\hline
$V^\mu_{-2/3} ~[U_1]$ & $- g_{\ell q}^{\mu b} (g_{\ell q}^{\mu s})^*$ & $ g_{\ell q}^{\mu b} (g_{\ell q}^{\mu s})^*$
& $- g_{e d}^{\mu b} (g_{e d}^{\mu s})^*$ & $- g_{e d}^{\mu b} (g_{e d}^{\mu s})^*$ \\
& $2 g_{\ell q}^{\mu b} (g_{e d}^{\mu s})^*$ & $2 (g_{\ell q}^{\mu s})^* g_{e d}^{\mu b}$ & 0 & 0 \\
\hline
${\vec V}^{\prime \mu}_{-2/3} ~[U_3]$ & $- g_{\ell q}^{\prime \mu b} (g_{\ell q}^{\prime \mu s})^*$
                                   & $ g_{\ell q}^{\prime \mu b} (g_{\ell q}^{\prime \mu s})^*$ & 0 & 0 \\
& 0 & 0 & $- 2 g_{\ell q}^{\prime \mu b} (g_{\ell q}^{\prime \mu s})^*$ & 0 \\
\hline
$V^\mu_{-5/6}$ & $g_{e q}^{\mu s} (g_{e q}^{\mu b})^*$ & $ g_{e q}^{\mu s} (g_{e q}^{\mu b})^*$
& $g_{\ell d}^{\mu s} (g_{\ell d}^{\mu b})^*$ & $- g_{\ell d}^{\mu s} (g_{\ell d}^{\mu b})^*$ \\
& $2 g_{\ell d}^{\mu s} (g_{e q}^{\mu b})^*$ & $2 (g_{\ell d}^{\mu b})^* g_{e q}^{\mu s}$ & 0 & $g_{\ell d}^{\mu s} (g_{\ell d}^{\mu b})^*$ \\
\hline
\end{tabular}
\end{center}
\caption{Contributions of the different LQs to the $\bsmumu$ WCs of
  various operators. Only the $V^\mu_{-2/3}$ and $V^\mu_{-5/6}$ LQs
  contribute to $O^{(\prime)}_{S,P}$, with $C_P^{\prime\mu\mu}({\rm
    NP}) = C_{S,\NP}^{\prime\mu\mu}$. The $\bsee$ WCs are obtained by
  changing $\mu \to e$ in the superscripts.  The normalization $K
  \equiv \pi / (\sqrt{2} \alpha G_F V_{tb} V_{ts}^* M_{LQ}^2)$ has
  been factored out. For $M_{LQ} = 1$ TeV, $K = -644.4$.
\label{LQWC}}
\end{table}

With this, we can answer the question of the introduction to this
section: can the scenarios in Tables \ref{NP_RK*lowq2_1} and
\ref{NP_RK*lowq2_2} be reproduced within LQ models? We see that all LQ
models have $C_{9,\NP} = \pm C_{10,\NP}$ and/or $C_{9,\NP}^{\prime} =
\pm C_{10,\NP}^{\prime}$ for both $\bsmumu$ and $\bsee$. However, for
the first four scenarios in Table \ref{NP_RK*lowq2_1}, these relations
do not hold, leading us to conclude that these solutions cannot be
reproduced with LQ models.

On the other hand, scenario S5 of Table \ref{NP_RK*lowq2_1} (which is
borderline) and the scenarios of Table \ref{NP_RK*lowq2_2} have no
unprimed-primed relations, so they can be explained with models
involving several different types of LQ. For example, consider
scenario S9 of Table \ref{NP_RK*lowq2_2}: $C_{9,\NP}^{\mu\mu} =
-C_{10,\NP}^{\mu\mu} = -0.52$, $C_{9,\NP}^{'ee} = 1.00$,
$C_{10,\NP}^{'ee} = 1.24$. One way to obtain this is to combine the
following LQs: ${\vec \Delta}'_{1/3}$ with $y_{\ell q}^{\prime \mu b}
(y_{\ell q}^{\prime \mu s})^* = -0.52$, $\Delta_{-1/6}$ with $\frac12
y_{\ell d}^{e b} (y_{\ell d}^{e s})^* = 0.12$, and $\Delta_{4/3}$ with
$\frac12 y_{e d}^{e b} (y_{e d}^{e s})^* = 1.12$. The other scenarios
can be reproduced with similar combinations of LQs. Note that, since
different LQs couple to $\mu$ and $e$, there are no contributions to,
and constraints from, LFV processes.

But this raises a modification of the question: using a model with a
single type of LQ, are there scenarios in which $R_{K^*}^{low}$ can be
explained with the addition of a contribution to $\bsee$? We begin
with the $\bsmumu$ WCs. As noted above, all LQ models have
$C_{9,\NP}^{\mu\mu} = \pm C_{10,\NP}^{\mu\mu}$ and/or
$C_{9,\NP}^{\prime\mu\mu} = \pm C_{10,\NP}^{\prime\mu\mu}$. However,
it has been shown that, of these four possibilities, the model must
include $C_{9,\NP}^{\mu\mu} = - C_{10,\NP}^{\mu\mu}$ to explain the
$\bsmumu$ data \cite{Alok:2017jgr}. This implies that only the $S_3$,
$U_1$ and $U_3$ LQ models are possible. Turning to the $\bsee$ WCs,
for $S_3$ and $U_3$ the only possibility is $C_{9,\NP}^{ee} = -
C_{10,\NP}^{ee}$, meaning that the LQ couplings involve only LH
particles. But scenario S7 of Table \ref{NP_RK*lowq2_1} shows that
this choice of NP WCs cannot explain $R_{K^*}^{low}$, so $S_3$ and
$U_3$ are excluded.

This leaves the $U_1$ LQ model as the only possibility. Its analysis
has the following ingredients:
\begin{itemize}

\item $\bsmumu$: The WCs for $U_1$ must include $C_{9,\NP}^{\mu\mu} =
  - C_{10,\NP}^{\mu\mu}$.  In principle, $C_{9,\NP}^{\prime\mu\mu} = +
  C_{10,\NP}^{\prime\mu\mu}$ could also be present. However, if these
  primed WCs are sizeable, so too are the scalar WCs
  $C_{S,\NP}^{\mu\mu}$ and $C_{S,\NP}^{\prime\mu\mu}$ (see Table
  \ref{LQWC}). The problem is that the scalar operators
  $O^{(\prime)\mu\mu}_S$ [Eq.~(\ref{newops})] contribute significantly
  to $\bs\to\mu^+\mu^-$ \cite{Alok:2010zd}, so that the present
  measurement of $\cB(\bs\to\mu^+\mu^-)$ \cite{Aaij:2013aka,
    CMS:2014xfa}, in agreement with the SM, puts severe constraints on
  $C_{S,\NP}^{(\prime)\mu\mu}$, and hence on $C_{9,\NP}^{\prime\mu\mu}
  = + C_{10,\NP}^{\prime\mu\mu}$.  For this reason, we keep only
  $C_{9,\NP}^{\mu\mu} = - C_{10,\NP}^{\mu\mu}$ as the nonzero
  $\bsmumu$ NP WCs.

\item $\bsee$: For the WCs, one can have $C_{9,\NP}^{ee} = -
  C_{10,\NP}^{ee}$, $C_{9,\NP}^{\prime ee} = C_{10,\NP}^{\prime ee}$,
  or both. The first case is excluded (see scenario S7 of Table
  \ref{NP_RK*lowq2_1}). The second case is allowed, but gives only a
  borderline result (see scenario S6 of Table \ref{NP_RK*lowq2_1}).
  This leaves the third case, with two independent combinations of WCs
  in $\bsee$.  As above, here the scalar operators $O^{(\prime) ee}_S$
  are generated, so the constraint $\cB(\bs\to e^+e^-) < 2.8 \times
  10^{-7}$ (90\% C.L.)  \cite{pdg2018} must be taken into account.
  Table~\ref{LQWC} shows that all $\bsee$ WCs can be written as
  functions of the four LQ couplings $g_{\ell q}^{e b}$, $g_{\ell
    q}^{e s}$, $g_{e d}^{e b}$ and $g_{e d}^{e b}$.

\item $b \to s \nu_\ell {\bar \nu}_{\ell^{(\prime)}}$: As can be seen
  in Table~\ref{LQWC}, the $U_1$ LQ model has $C_{\nu,\NP}^{(\prime)
    \mu\mu} = 0$, so there are no additional constraints from $b \to s
  \nu_\ell {\bar \nu}_{\ell^{(\prime)}}$.

\item LFV processes:
\begin{itemize}

\item $b\to s e^+ \mu^-$: The nonzero WCs are
\beq
C_{9,\NP}^{e\mu} = -C_{10,\NP}^{e\mu}= -g_{\ell q}^{\mu b}  (g_{\ell q}^{e s} )^* ~~,~~~~
C_{S,\NP}^{e\mu} =  2 g_{\ell q}^{\mu b}  (g_{ed}^{e s} )^* ~.
\eeq

\item $b\to s \mu^+ e^-$: The nonzero WCs are
\beq
C_{9,\NP}^{\mu e} = -C_{10,\NP}^{\mu e}= -g_{\ell q}^{e b}  (g_{\ell q}^{\mu s} )^* ~~,~~~~
C_{S,\NP}^{\prime,\mu e} =  2 (g_{\ell q}^{\mu s})^*  g_{ed}^{e b} ~.
\eeq

\item $\mu\to e\gamma$: The WCs are \cite{Crivellin:2017dsk}
\beq
C_{\gamma}^{L} = \frac{e N_c m_\mu}{16 \pi^2 M_{LQ}^2} \,
\frac16 ( g_{\ell q}^{eb} g_{\ell q}^{\mu b} + g_{\ell q}^{es} g_{\ell q}^{\mu s} ) ~~,~~~~ C_{\gamma}^R = 0 ~.
\eeq

\end{itemize}
The experimental measurements of the LFV observables are given in
Table \ref{tab:lfv}.

\begin{table*}[t]
\renewcommand{\arraystretch}{2}
\begin{center}
\begin{tabular}{|c|c|c|}
\hline
Observables &  Measurement\\ \hline
$\mathcal{B}(B^+\to K^+ \mu^+ e^-)$&  $(-12.1^{+7.4}_{-5.0}\pm 2.3)\times 10^{-8}$  \cite{Aubert:2006vb}\\
$\mathcal{B}(B^+\to K^+ \mu^- e^+)$&  $(-2.9^{+7.4}_{-4.4}\pm 1.9)\times 10^{-8}$  \cite{Aubert:2006vb}\\
$\mathcal{B}(B\to K^* \mu^- e^+)$  & $(7.0^{+23}_{-13} \pm 5)\times 10^{-8}$  \cite{Aubert:2006vb}\\
$\mathcal{B}(B\to K^* \mu^+ e^-)$  & $(-7.0^{+22}_{-14} \pm 7)\times 10^{-8}$  \cite{Aubert:2006vb}\\
$\mathcal{B}(B^+\to K^{*+} \mu^- e^+)$ & $(9.0^{+65}_{-44} \pm 22)\times 10^{-8}$  \cite{Aubert:2006vb}\\
$\mathcal{B}(B^+\to K^{*+} \mu^+ e^-)$ & $(-32^{+63}_{-38} \pm 15)\times 10^{-8}$  \cite{Aubert:2006vb}\\
$\mathcal{B}(B_s\to \mu^\pm e^\mp)$   & $< 6.3 \times 10^{-9}$ ~~(95\% C.L.) \cite{Aaij:2017cza}\\
$\mathcal{B}(\mu \to e \gamma)$    &  $<4.2 \times 10^{-13}$  ~~(90\% C.L.) \cite{pdg2018} \\
\hline
\end{tabular}
\end{center}
\caption{Measurements of LFV observables.}
\label{tab:lfv}
\end{table*}

\end{itemize}

The analysis of the $U_1$ LQ therefore involves a fit with six unknown
parameters: $g_{\ell q}^{\mu b} $, $g_{\ell q}^{\mu s} $, $g_{\ell
  q}^{e b}$, $g_{\ell q}^{e s}$, $g_{e d}^{e b}$ and $g_{e d}^{e s}$.
We fix $C_{9,\NP}^{\mu\mu} = - C_{10,\NP}^{\mu\mu} = 644.4 \, g_{\ell
  q}^{\mu b} (g_{\ell q}^{\mu s})^*$ to its central value, $-0.62$
[Eq.~(\ref{bsmumuNP})]. For simplicity, we assume that all couplings
are real and take $g_{\ell q}^{\mu b} = -g_{\ell q}^{\mu s} =
0.03$. The best-fit values and (correlated) errors of the four unknown
couplings are found to be
\beq
g_{\ell q}^{e b} = -0.01 \pm 0.05 ~,~~
g_{\ell q}^{e s} = -0.007 \pm 0.030 ~,~~
g_{e d}^{e b} = 0.003 \pm 0.002 ~,~~
g_{e d}^{e s} = 3.0 \times 10^{-4} \pm 0.02 ~.
\eeq
The LFV constraints are clearly very stringent, as the central values
of the couplings are all very near zero. The errors are larger, but,
even so, when the couplings are varied within their 68\% C.L.-allowed
region, the smallest predicted value of $R_{K^*}^{low}$ is 0.82, which
is quite a bit larger than $1\sigma$ above its measured value. If
different values of $g_{\ell q}^{\mu b}$ and $g_{\ell q}^{\mu s}$ are
chosen, all the while satisfying $644.4 \, g_{\ell q}^{\mu b} (g_{\ell
  q}^{\mu s})^* = -0.62$, the best-fit values and errors of the
couplings are of course different. However, we have verified that the
prediction for $R_{K^*}^{low}$ does not improve.

We therefore conclude that the experimental result for $R_{K^*}^{low}$
cannot be explained within the $U_1$ LQ model alone. More generally,
this result cannot be explained using a model with a single type of
LQ.

\subsubsection{$Z'$ gauge bosons}

A $Z'$ is typically the gauge boson associated with an additional
$U(1)'$. As such, in the most general case, it has independent
couplings to the various pairs of fermions. As we are focused on
$\bsmumu$ and $\bsee$ transitions, the couplings that interest us are
$g_L^{sb}$, $g_R^{sb}$, $g_L^\mu$, $g_R^\mu$, $g_L^e$ and $g_R^e$,
which are the coefficients of $({\bar s} \gamma^\mu P_L b)Z'_\mu$,
$({\bar s} \gamma^\mu P_R b)Z'_\mu$, $({\bar \mu} \gamma^\mu P_L
\mu)Z'_\mu$, $({\bar \mu} \gamma^\mu P_R \mu)Z'_\mu$, $({\bar e}
\gamma^\mu P_L e)Z'_\mu$ and $({\bar e} \gamma^\mu P_R e)Z'_\mu$,
respectively. We define $g_V^\ell \equiv g_R^\ell + g_L^\ell$ and
$g_A^\ell \equiv g_R^\ell - g_L^\ell$ $(\ell = \mu, e)$. We can then
write
\bea
& C_{9,\NP}^{\mu\mu} = K \, g_L^{sb} g_V^\mu ~,~~
C_{10,\NP}^{\mu\mu} = K \, g_L^{sb} g_A^\mu ~,~~
C_{9,\NP}^{\prime\mu\mu} = K \, g_R^{sb} g_V^\mu ~,~~
C_{10,\NP}^{\prime\mu\mu} = K \, g_R^{sb} g_A^\mu ~, & \nn\\
& C_{9,\NP}^{ee} = K \, g_L^{sb} g_V^e ~,~~
C_{10,\NP}^{ee} = K \, g_L^{sb} g_A^e ~,~~
C_{9,\NP}^{\prime ee} = K \, g_R^{sb} g_V^e ~,~~
C_{10,\NP}^{\prime ee} = K \, g_R^{sb} g_A^e ~, &
\eea
where
\beq
K \equiv \pi / (\sqrt{2} \alpha G_F V_{tb} V_{ts}^* M_{Z'}^2) = -644.4 ~~({\rm for}~M_{Z'} = 1~{\rm TeV}) ~.
\eeq
Given that there are six couplings and eight WCs, there must be
relations among the WCs. They are
\beq
\frac{C_{9,\NP}^{\mu\mu}}{C_{9,\NP}^{\prime\mu\mu}} =
\frac{C_{10,\NP}^{\mu\mu}}{C_{10,\NP}^{\prime\mu\mu}} =
\frac{C_{9,\NP}^{ee}}{C_{9,\NP}^{\prime ee}} =
\frac{C_{10,\NP}^{ee}}{C_{10,\NP}^{\prime ee}} ~.
\label{WCrelations}
\eeq

In general, other processes may be affected by $Z'$ exchange, and
these produce constraints on the couplings. One example is
$\bs$-$\bsbar$ mixing: since the $Z'$ couples to ${\bar s}b$, there is
a tree-level contribution to this mixing. When the $Z'$ is integrated
out, one obtains the four-fermion operators
\beq
\frac{(g_L^{sb})^2}{2 M^2_{Z'}} \, ({\bar s}_L \gamma^\mu b_L)\,({\bar s}_L \gamma_\mu b_L) 
+ \frac{(g_R^{sb})^2}{2 M^2_{Z'}} \, ({\bar s}_R \gamma^\mu b_R)\,({\bar s}_R \gamma_\mu b_R) 
+ \frac{g_L^{sb} g_R^{sb}}{M^2_{Z'}} \, ({\bar s}_L \gamma^\mu b_L)\,({\bar s}_R \gamma_\mu b_R) ~,
\label{BsmixingVB}
\eeq
all of which contribute to $\bs$-$\bsbar$ mixing. We refer to these as
the $LL$, $RR$ and $LR$ contributions, respectively.  The $LL$ term
has been analyzed most recently in Ref.~\cite{Kumar:2018kmr}. There it
is found that the comparison of the measured value of $\bs$-$\bsbar$
mixing with the SM prediction implies
\beq
\frac{g_L^{sb}}{M_{Z'}} = \pm (1.0^{+2.0}_{-3.9}) \times 10^{-3}~{\rm TeV}^{-1} ~.
\label{EQ:VB:Bsmixing}
\eeq
The $RR$ term yields a similar constraint on $g_R^{sb}$. The $LR$
contribution has been examined in Ref.~\cite{Crivellin:2015era} -- the
constraint one obtains on $g_L^{sb} g_R^{sb}$ is satisfied once one
imposes the above individual constraints on $g_L^{sb}$ and $g_R^{sb}$.
(We note in passing that the model in Ref.~\cite{Guadagnoli:2018ojc}
is constructed such that all contributions to $\bs$-$\bsbar$ mixing
vanish.)

The coupling of the $Z'$ to $\mu^+\mu^-$ can be constrained by the
measurement of the production of $\mu^+\mu^-$ pairs in
neutrino-nucleus scattering, $\nu_\mu N \to \nu_\mu N \mu^+ \mu^-$
(neutrino trident production). Ref.~\cite{Kumar:2018kmr} finds
\beq
\frac{g_L^{\mu\mu}}{M_{Z'}} = 0 \pm 1.13~{\rm TeV}^{-1} ~.
 \label{EQ:VB:trident}
\eeq
The constraint on $g_R^{\mu\mu}$ is much weaker, since it does not
interfere with the SM. Note that, with $g_{L,R}^{sb} \lsim O(10^{-3})$
and $g_{L.R}^{\mu\mu} = O(1)$, the expected sizes of the $\bsmumu$ NP
WCs are $C_{9,10,\NP}^{(\prime)\mu\mu} \lsim 0.6$, which is what is
found in the various scenarios.

With the relations in Eq.~(\ref{WCrelations}), it is straightforward
to verify that the first four scenarios in Table \ref{NP_RK*lowq2_1}
cannot be reproduced with the addition of a $Z'$. For example, in
scenario S1 of the Table, $C_{9,10,\NP}^{\prime\mu\mu} = 0$, which can
occur only if $g_R^{sb} = 0$. This then implies $C_{10,\NP}^{\prime
  ee} = 0$, in contradiction with the nonzero value of
$C_{10,\NP}^{\prime ee}$ required in this scenario. A similar logic
applies to solutions S2, S3 and S4 in Table \ref{NP_RK*lowq2_1}. On
the other hand, scenario S5, which is borderline, {\it can} be
produced within a $Z'$ model -- all that is required is that
$g_R^{sb}$, $g_R^\mu$ and $g_L^e$ vanish.

Turning to Table \ref{NP_RK*lowq2_2}, scenarios S9 and S11 cannot be
explained by a $Z'$ model for the same reason. On the other hand, the
addition of a $Z'$ {\it can} reproduce scenarios S8 and S10, which
involve only unprimed WCs.

Finally, we consider more general scenarios involving all eight WCs,
taking into account the relations in Eq.~(\ref{WCrelations}). With six
independent couplings, there are a great many possibilities to
consider. We first try $1 + 1$ scenarios:
\bea
& (1a) & g_L^{sb} = g_R^{sb} ~,~~ g_V^\mu = -g_A^\mu ~,~~ g_V^e = -g_A^e \nn\\
&& \Longrightarrow
C_{9,\NP}^{\mu\mu} = -C_{10,\NP}^{\mu\mu} = C_{9,\NP}^{\prime\mu\mu} = -C_{10,\NP}^{\prime\mu\mu} ~,~~
C_{9,\NP}^{ee} = -C_{10,\NP}^{ee} = C_{9,\NP}^{\prime ee} = -C_{10,\NP}^{\prime ee} ~, \nn\\
& (1b) & g_L^{sb} = -g_R^{sb} ~,~~ g_V^\mu = -g_A^\mu ~,~~ g_V^e = -g_A^e \\
&& \Longrightarrow 
C_{9,\NP}^{\mu\mu} = -C_{10,\NP}^{\mu\mu} = -C_{9,\NP}^{\prime\mu\mu} = C_{10,\NP}^{\prime\mu\mu} ~,~~
C_{9,\NP}^{ee} = -C_{10,\NP}^{ee} = -C_{9,\NP}^{\prime ee} = C_{10,\NP}^{\prime ee} ~. \nn
\eea
However, neither of these gives a good fit to the data. This is due to
the $\bsmumu$ NP WCs: it is well known that, in order to explain the
data, the NP must be mainly in $C_{9,10,\NP}^{\mu\mu}$, which have a
left-handed coupling to the quarks \cite{Descotes-Genon:2015uva}. The
right-handed NP WCs $C_{9,10,\NP}^{\prime\mu\mu}$ may be nonzero, but
they must be smaller than $C_{9,10,\NP}^{\mu\mu}$, which is not the
case above.

In light of this, we try the following $2 + 2$ scenarios:
\bea
& (2a) & g_L^{sb}, g_R^{sb}~{\rm free} ~,~~ g_V^\mu = -g_A^\mu ~,~~ g_V^e = -g_A^e \nn\\
&& \Longrightarrow
C_{9,\NP}^{\mu\mu} = -C_{10,\NP}^{\mu\mu} ~,~~ C_{9,\NP}^{\prime\mu\mu} = -C_{10,\NP}^{\prime\mu\mu} ~,~~
C_{9,\NP}^{ee} = -C_{10,\NP}^{ee} ~,~~ C_{9,\NP}^{\prime ee} = -C_{10,\NP}^{\prime ee} ~, \nn\\
& (2b) & g_L^{sb}, g_R^{sb}~{\rm free} ~,~~ g_V^\mu = -g_A^\mu ~,~~ g_V^e = g_A^e \\
&& \Longrightarrow 
C_{9,\NP}^{\mu\mu} = -C_{10,\NP}^{\mu\mu} ~,~~ C_{9,\NP}^{\prime\mu\mu} = -C_{10,\NP}^{\prime\mu\mu} ~,~~
C_{9,\NP}^{ee} = C_{10,\NP}^{ee} ~,~~ C_{9,\NP}^{\prime ee} = C_{10,\NP}^{\prime ee} ~. \nn
\eea
For both of these cases, we find that a value for $R_{K^*}^{low}$ is
predicted within roughly $1\sigma$ of its measured value. The details
are shown in Table \ref{Z'solution}.

\begin{table*}[t] 
\renewcommand{\arraystretch}{2}
\begin{center}
\begin{tabular}{|c|l|l|c|c|c|c|}
 \hline 
& NP in $\bsmumu$ & NP in $\bsee$ & $R_{K^*}^{low}$ & $R_{K^*}^{cen}$ & $R_K$ & Pull \\
\hline
S12 & $C_{9,\NP}^{\mu\mu} = -C_{10,\NP}^{\mu\mu}$ & $C_{9,\NP}^{ee} = -C_{10,\NP}^{ee}$ & & & & \\
& ~~~ $= -0.61 \pm 0.11$ & ~~~ $= 0.08 \pm 0.20$ & & & & \\
& $C_{9,\NP}^{\prime \mu\mu} = -C_{10,\NP}^{\prime \mu\mu}$ & $C_{9,\NP}^{\prime ee} = -C_{10,\NP}^{\prime ee}$ & & & & \\
& ~~ $= 0.16 \pm 0.09$ & ~~~ $= -0.03 \pm 0.20$ & (0.76) 0.82 & (0.53) 0.65 & (0.79) 0.77 & 6.6 \\
\hline
S13 & $C_{9,\NP}^{\mu\mu} = -C_{10,\NP}^{\mu\mu}$ & $C_{9,\NP}^{ee} = C_{10,\NP}^{ee}$ & & & & \\
& ~~~ $= -0.69 \pm 0.12$ & ~~~ $= -0.20 \pm 0.69$ & & & & \\
& $C_{9,\NP}^{\prime \mu\mu} = -C_{10,\NP}^{\prime \mu\mu}$ & $C_{9,\NP}^{\prime ee} = C_{10,\NP}^{\prime ee}$ & & & & \\
& ~~~ $= 0.14 \pm 0.08$ & ~~~ $= 0.14 \pm 0.97$ & (0.76) 0.82 & (0.52) 0.63 & (0.75) 0.74 & 6.5 \\
\hline
\end{tabular}
\end{center}
\caption{$Z'$-model scenarios that can generate a value for
  $R_{K^*}^{low}$ within $1\sigma$ of its measured value. Predictions
  for $R_{K^*}^{cen}$, $R_K$ and the pull are also shown.}
\label{Z'solution}
\end{table*}

\section{\boldmath Effects of New Physics in $\bsee$}

\subsection{$R_{K^{(*)}}$ Predictions}

In the introduction it was noted that NP in $\bsee$ is independent of
$q^2$. That is, the effect on $R_K$ should be the same, regardless of
whether $0.045 \le q^2 \le 1.1 ~{\rm GeV}^2$ (low), $1.1 \le q^2 \le
6.0 ~{\rm GeV}^2$ (central) or $15.0 \le q^2 \le 19.0 ~{\rm GeV}^2$
(high), and similarly for $R_{K^*}$. In fact, this is not completely
true. At low $q^2$, the $m_\mu - m_e$ mass difference is important for
$R_{K^*}$ (which is why the SM predicts $R_{K^*}^{low} \simeq 0.93$,
but $R_{K^*}^{cen,high} = 1$ \cite{flavio}). In addition, photon
exchange plays a more important role at low $q^2$ than in higher $q^2$
bins. As a result the correction due to NP in $\bsee$ will be
different for $R_{K^*}^{low}$ than it is for $R_{K^*}^{cen,high}$.
However, this does not apply to $R_K$ -- the NP effects are the same
for all $q^2$ bins.

To see this explicitly, below we present the numerical expressions for
$R_{K^{(*)}}$ as linearized functions of the WCs. These are obtained
using flavio \cite{flavio}.
\bea
R_{K^*}^{low} &\simeq&  0.93 +  0.04~ \left (C_{9,NP}^{\mu\mu}-C_{9,NP}^{ee} \right)  -0.09\left(C_{10,NP}^{\mu\mu}-C_{10,NP}^{ee}\right) \nn\\
&& -~0.07 \left(C_{9',NP}^{\mu\mu}-C_{9',NP}^{ee}\right) + 0.08\left(C_{10',NP}^{\mu\mu}- C_{10',NP}^{ee} \right) ~, \nn\\
R_{K^*}^{cen,high} &\simeq& 1.0 + 0.18 \left(C_{9,NP}^{\mu\mu}-C_{9,NP}^{ee}\right)  -0.29\left(C_{10,NP}^{\mu\mu}-C_{10,NP}^{ee}\right) \nn\\
&& -~0.19\left(C_{9',NP}^{\mu\mu}-C_{9',NP}^{ee}\right) + 0.22\left(C_{10',NP}^{\mu\mu} -C_{10',NP}^{ee}\right) ~, \nn\\
R_{K}^{low,cen,high} &\simeq& 1.0 + 0.24\left(C_{9,NP}^{\mu\mu}- C_{9,NP}^{ee}\right)  -0.26\left(C_{10,NP}^{\mu\mu}-C_{10,NP}^{ee}\right) \nn\\
&& +~0.24\left(C_{9',NP}^{\mu\mu}-C_{9',NP}^{ee}\right) - 0.26\left(C_{10',NP}^{\mu\mu} -C_{10',NP}^{ee}\right) ~.
\eea
We see that the expression for $R_{K^*}^{low}$ is different from that
for $R_{K^*}^{cen,high}$. The coefficients of the various terms are
larger in $R_{K^*}^{cen,high}$ than in $R_{K^*}^{low}$. Still, they
have the same signs, suggesting that the effect of NP in $\bsee$ is to
lower (or increase) the values of both $R_{K^*}^{low}$ and
$R_{K^*}^{cen,high}$. (However, since there are several terms, of
differing signs, this need not always be the case.) For $R_K$, the
expressions are essentially the same for the low, central and high
ranges of $q^2$. And since some of the coefficients of the various
terms in $R_{K}^{low,cen,high}$ have different signs than in
$R_{K^*}^{low,cen,high}$, the effect on $R_K$ of NP in $\bsee$ is
uncorrelated with its effect on $R_{K^*}$.

This is then a prediction. If the small experimental measured value of
$R_{K^*}^{low}$ is due to the presence of NP in $\bsee$, we expect
that future measurements will find $R_{K^*}^{cen} = R_{K^*}^{high}$
and $R_{K}^{low} = R_{K}^{cen} = R_{K}^{high}$. (This is a generic
prediction of any $q^2$-independent NP.)

\subsection{$Q_{4,5}$ Predictions}

$R_K$ and $R_{K^*}$ are Lepton-Flavour-Universality-Violating (LFUV)
observables. Any explanation of their measured values can be tested by
measuring other LFUV observables, such as $Q_i \equiv
P^{\prime\mu\mu}_i - P^{\prime ee}_i$ ($i=4,5$). Here,
$P^{\prime\ell\ell}_i$ are extracted from the angular distribution of
$B \to K^* \ell^+ \ell^-$. $Q_{4,5}$ have been measured at Belle
\cite{Wehle:2016yoi}. The results for $1.0 \le q^2 \le 6.0 ~{\rm
  GeV}^2$ are
\beq
Q_4 = 0.498 \pm 0.527 \pm 0.166 ~~,~~~~
Q_5 = 0.656 \pm 0.485 \pm 0.103 ~.
\label{Q45meas}
\eeq
At present, the errors are still very large.

The numerical expressions for these quantities as linearized functions
of the WCs are \cite{flavio}
\bea
Q_{4} &\simeq&  -0.03\left(C_{9,NP}^{\mu\mu}- C_{9,NP}^{ee} \right) +0.05\left(C_{10,NP}^{\mu\mu}-C_{10,NP}^{ee}\right) \nn\\
&& +~0.03\left(C_{9',NP}^{\mu\mu}-C_{9',NP}^{ee}\right) - 0.11\left(C_{10',NP}^{\mu\mu} -C_{10',NP}^{ee}\right) ~, \nn\\
Q_{5} &\simeq&   -0.24\left(C_{9,NP}^{\mu\mu}-C_{9,NP}^{ee} \right) -0.03\left(C_{10,NP}^{\mu\mu}-C_{10,NP}^{ee}\right) \nn \\
&& -~0.06\left(C_{9',NP}^{\mu\mu}-C_{9',NP}^{ee} \right)+ 0.22\left(C_{10',NP}^{\mu\mu}-C_{10',NP}^{ee}\right) ~. 
\eea
The coefficients of the various terms are generally larger in $Q_5$
than in $Q_4$, suggesting that the NP effect on $Q_5$ will be more
important. 

\begin{figure}[h]
\begin{center}
\includegraphics[width=0.6\textwidth]{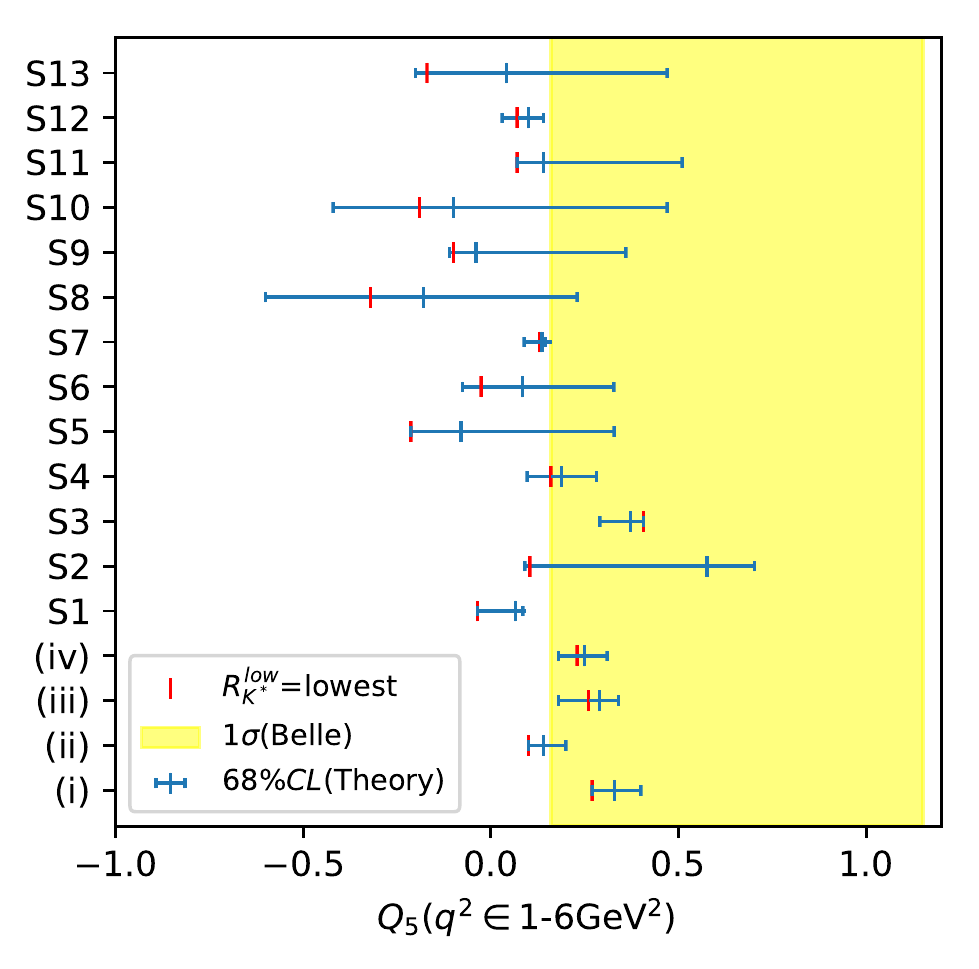}
\caption{Predicted range of values of $Q_5$ for each of the scenarios
  in Tables \ref{NP_RK*lowq2_1}, \ref{NP_RK*lowq2_2} and
  \ref{Z'solution}, as well as scenarios (i), (ii), (iii) and (iv)
      [Eqs.~(\ref{bsmumuNP}), (\ref{scenario(iii)}) and
        (\ref{scenario(iv)})]. The $1\sigma$ range of the present
      measurement of $Q_5$ [Eq.~(\ref{Q45meas})] is superposed.}
\label{Q5predictions}
\end{center}
\end{figure}

Indeed, a future precise measurement of $Q_5$ will give us a great
deal of information. In Fig.~\ref{Q5predictions} we present the
predictions for $Q_5$ of the various scenarios described in Tables
\ref{NP_RK*lowq2_1}, \ref{NP_RK*lowq2_2} and \ref{Z'solution}, as well
as scenarios (i), (ii), (iii) and (iv) [Eqs.~(\ref{bsmumuNP}),
  (\ref{scenario(iii)}) and (\ref{scenario(iv)})]. We superpose the
present Belle measurement [Eq.~(\ref{Q45meas})]. We see the following:
\begin{itemize}

\item Certain scenarios (e.g., S2, S8, S10, S13) predict a rather wide
  range of values of $Q_5$.  However, for the other scenarios, the
  predicted range is fairly small, so that, if $Q_5$ is measured
  reasonably precisely, we will be able to exclude some of them. In
  other words, a good measurement of $Q_5$ will provide an important
  constraint on scenarios constructed to explain $R_{K^*}^{low}$ via
  the addition of NP in $\bsee$.

\item If there is NP only in $\bsmumu$ [scenarios (i), (ii), (iii) and
  (iv)], $Q_5$ is predicted to be positive. This is due to the fact
  that, in all four scenarios, $C_{9,\NP}^{\mu\mu}$ is large and
  negative. If $Q_5$ were found to be negative, this would be a clear
  signal that NP only in $\bsmumu$ is insufficient. And indeed,
  several scenarios with NP in $\bsee$ allow for $Q_5 < 0$ within
  their 68\% C.L.\ ranges.

\end{itemize}

\subsection{LFUV and LFU New Physics}

As noted above, $R_K$ and $R_{K^*}$ are LFUV observables. On the other
hand, the processes $B \to K^* \mu^+\mu^-$ and $B_s^0 \to \phi \mu^+
\mu^-$ are governed by $\bsmumu$ transitions. The associated
observables are Lepton-Flavour Dependent (LFD). In order to explain
the anomalies in $B$ decays, most analyses have assumed NP only in
$\bsmumu$, i.e., purely LFUV NP. Recently, in
Ref.~\cite{Alguero:2018nvb}, it is suggested to modify the NP paradigm
by considering in addition Lepton-Flavour-Universal (LFU) NP. The LFUV
observables are then explained by the LFUV NP, while the LFD
observables are explained by LFUV $+$ LFU NP. Numerous scenarios are
constructed with both LFUV and LFU NP that explain the data as well as
scenarios with only LFUV NP.

In Ref.~\cite{Alguero:2018nvb}, the addition of LFU NP was not a
necessity, but was seen as a logical possibility.  In the present
paper, we add NP in $\bsee$ specifically with the aim of improving the
explanation of the measured value of $R_{K^*}^{low}$. Technically,
this is not LFU NP, but it can be made so by including equal WCs in $b
\to s \tau^+ \tau^-$ transitions. All our scenarios can be translated
into LFUV $+$ LFU NP. Conversely, the scenarios of
Ref.~\cite{Alguero:2018nvb} can be translated into $\bsmumu$ NP $+$
$\bsee$ NP. As such, the two papers are complementary to one another.

Here is an example. Ref.~\cite{Alguero:2018nvb} performs the analysis
in terms of the LFUV WCs $C^V_{i\ell}$ and the LFU WCs $C_i^U$ ($i =
9,10$, $\ell = e, \mu$). Without loss of generality, they set
$C^V_{ie} = 0$. In the most general case, where all four WCs are free,
the best-fit values of the WCs are found to be
\beq
C^V_{9\mu} = 0.08 ~,~~ C^V_{10\mu} = 1.14 ~,~~ C^U_9 = -1.26 ~,~~ C^U_{10} = -0.91 ~.
\eeq
Converting these to $\bsmumu$ and $\bsee$ WCs, one obtains
\beq
C_{9,\NP}^{\mu\mu} = -1.18 ~,~~ C_{10,\NP}^{\mu\mu} = 0.23 ~,~~ C_{9,\NP}^{ee} = -1.26 ~,~~ C_{10,\NP}^{ee} = -0.91 ~.
\eeq
These are to be compared with the best-fit values of the WCs in
scenario S10 of Table \ref{NP_RK*lowq2_2}.  The agreement is
excellent. We therefore see that our scenario S10 is equivalent to
the most general LFUV/LFU scenario of Ref.~\cite{Alguero:2018nvb}.
That is, this LFUV/LFU scenario can explain the measured value of
$R_{K^*}^{low}$.

Now, we have found a number of other scenarios which can account for
$R_{K^*}^{low}$.  However, they involve the WCs $C_{9,\NP}^{\prime ee}$
and/or $C_{10,\NP}^{\prime ee}$. In Ref.~\cite{Alguero:2018nvb},
the focus was on LFUV NP only in $C_{9,10,\NP}^{\mu\mu}$. We have
given a motivation for also considering LFUV NP in
$C_{9,10,\NP}^{\prime ee}$. Indeed, from a mdel-building point of
view, it is quite natural to have both unprimed and primed NP WCs.

\section{Conclusions}

There are presently disagreements with the predictions of the SM in
the measurements of several observables in $B \to K^* \mu^+\mu^-$ and
$B_s^0 \to \phi \mu^+ \mu^-$ decays, and in the LFUV ratios $R_K$ and
$R_{K^*}$. Combining the various $B$ anomalies, analyses find that the
net discrepancy with the SM is at the level of 4-6$\sigma$. It is also
shown that, by adding NP only to $\bsmumu$, one can get a good fit to
the data. However, not all discrepancies are explained: there is still
a disagreement of $\gsim 1.7 \sigma$ with the measured value of
$R_{K^*}$ at low values of $q^2$. Of course, from the point of view of
a global fit, this disagreement is not important. Still, it raises the
question: if the true value of $R_{K^*}^{low}$ is near its measured
value, what can explain it?

If there is NP in $\bsmumu$, it would not be at all surprising if
there were also NP in $\bsmumu$.  In this paper, we show that, if NP
in $\bsee$ transitions is also allowed, one can generate values for
$R_{K^*}^{low}$ within $\sim 1\sigma$ of its measured value. We have
constructed a number of different scenarios (i.e., sets of $\bsmumu$
and $\bsee$ Wilson coefficients) in which this occurs. Some have one
NP WC (or combination of WCs) in each of $\bsmumu$ and $\bsee$, and
some have more NP WCs (or combinations of WCs) in $\bsmumu$ and/or
$\bsee$.

The analysis is done in part using a model-independent,
effective-field-theory approach. When one has NP only in $\bsmumu$, a
popular choice is $C_{9,\NP}^{\mu\mu} = -C_{10,\NP}^{\mu\mu}$, i.e.,
purely LH NP couplings. We find that, if the NP couplings in $\bsee$
are also purely LH, i.e., $C_{9,\NP}^{ee} = -C_{10,\NP}^{ee}$,
$R_{K^*}^{low}$ can {\it not} be explained.  $\bsee$ NP couplings
involving the RH quarks and/or leptons must be involved. 

With NP in both $\bsmumu$ and $\bsee$, one has a better agreement with
the data, leading to a bigger pull with respect to the SM. Even so, to
get a prediction for $R_{K^*}^{low}$ within $\sim 1\sigma$ of its
measured value, one has to use $\bsee$ WCs that are not the best-fit
values, but rather lie elsewhere within the 68\% C.L. region. At the
level of the goodness-of-fit, this costs very little: the pull is
reduced only by ~0.2 (i.e., a few percent).

We also perform the analysis using specific models. We find that, with
the addition of $\bsee$ NP couplings, the measured value of
$R_{K^*}^{low}$ can be explained within a model that includes several
different types of leptoquark, or with a model containing a $Z'$ gauge
boson.

Finally, NP in $\bsee$ is independent of $q^2$. For each scenario, we
can predict the values of $R_{K^*}$ and $R_K$ to be found in other
$q^2$ bins. We also show that a future precise measurement of $Q_5
\equiv P^{\prime\mu\mu}_5 - P^{\prime ee}_5$ will help in
distinguishing the various scenarios. It can also distinguish
scenarios with NP only in $\bsmumu$ from those in which NP in $\bsee$
is also present.

\bigskip
\noindent
{\bf Note added}: recently, at Moriond 2019, LHCb presented new $R_K$
results \cite{LHCbRKnew} and Belle presented its measurement of
$R_{K^*}$ \cite{BelleRK*new}. Following these announcements, global
fits using the new data were performed in Refs.~\cite{Alguero:2019ptt,
  Alok:2019ufo, Ciuchini:2019usw, Datta:2019zca, Aebischer:2019mlg,
  Kowalska:2019ley}, and it was found that the discrepancy with the
predictions of the SM is still sizeable. In three of these studies
\cite{Alguero:2019ptt, Datta:2019zca, Aebischer:2019mlg}, separate
fits to the $\bsmumu$ and $R_{K^{(*)}}$ data were performed. The
result was that there is now a tension between these two fits: under
the assumption that NP enters only in $\bsmumu$, the best-fit values
of the NP WCs differ by $\gsim 1\sigma$. This tension can be removed
by also allowing for NP in $\bsee$. In Ref.~\cite{Datta:2019zca}, the
additional NP contributions appear only in $\bsee$, while in
Refs.~\cite{Alguero:2019ptt, Aebischer:2019mlg},
lepton-flavour-universal NP contributions to both $\bsmumu$ and
$\bsee$ are added.

\bigskip
\noindent
{\bf Acknowledgments}: This work was financially supported in part by
NSERC of Canada.

\end{document}